\documentclass[conference]{IEEEtran}

\usepackage{varwidth}
\usepackage{hyperref}
\usepackage{algorithm}
\usepackage{algpseudocode}
\usepackage{graphicx}
\usepackage{tabularx}
\usepackage{subfig}
\usepackage[simplified]{pgf-umlcd}
\usepackage{tikz}
\usepackage{amsmath}
\usepackage{multirow}
\usepackage{array}
\usepackage{todonotes}
\usetikzlibrary{spy,arrows,fit,shapes,calc,snakes,trees,decorations.pathmorphing}
\usepackage{smartdiagram}

\begin{document}
\title{Dependency-Aware Rollback and Checkpoint-Restart for Distributed Task-Based Runtimes}

\author{
\IEEEauthorblockN{Kiril Dichev}
\IEEEauthorblockA{
%School of EEECS\\
Queen's University Belfast\\
Email: K.Dichev@qub.ac.uk}
\and
\IEEEauthorblockN{Herbert Jordan}
\IEEEauthorblockA{
University of Innsbruck\\
Herbert.Jordan@uibk.ac.at}
\and
%%\IEEEauthorblockN{Khalid Hasanov}
%%\IEEEauthorblockA{
%%IBM\\
%%blob@blob.com}
%\and
\IEEEauthorblockN{Konstantinos Tovletoglou}
\IEEEauthorblockA{
%School of EEECS\\
Queen's University Belfast\\
ktovletoglou01@qub.ac.uk}
\and
\IEEEauthorblockN{Thomas Heller}
\IEEEauthorblockA{
Friedrich-Alexander University Erlangen-Nuremberg\\
thomas.heller@cs.fau.de}
\and
\IEEEauthorblockN{Dimitrios~S.  Nikolopoulos}
\IEEEauthorblockA{
%School of EEECS\\
Queen's University Belfast\\
Email: D.Nikolopoulos@qub.ac.uk}
\and
\IEEEauthorblockN{Georgios Karakonstantis}
\IEEEauthorblockA{
Queen's University Belfast\\
G.Karakonstantis@qub.ac.uk}
\and
\IEEEauthorblockN{Charles Gillan}
\IEEEauthorblockA{
Queen's University Belfast\\
C.Gillan@qub.ac.uk}
}
\maketitle

\newtheorem{post}{Postulate}
\begin{abstract}
    With the increase in compute nodes in large compute platforms, a proportional increase in node failures will follow. Many application-based checkpoint/restart (C/R) techniques have been proposed for MPI applications to target the reduced mean time between failures. However, rollback as part of the recovery remains a dominant cost even in highly optimised MPI applications employing C/R techniques. Continuing execution past a checkpoint (that is, reducing rollback) is possible in message-passing runtimes, but extremely complex to design and implement. Our work focuses on task-based runtimes, where task dependencies are explicit and message passing is implicit. We see an opportunity for reducing rollback for such runtimes: we explore task dependencies in the rollback, which we call dependency-aware rollback. We also design a new C/R technique, which is influenced by recursive decomposition of tasks, and combine it with dependency-aware rollback. We expect the dependency-aware rollback to cancel and recompute less tasks in the presence of node failures. We describe, implement and validate the proposed protocol in a simulator, which confirms these expectations. In addition, we consistently observe faster overall execution time for dependency-aware rollback in the presence of faults, despite the fact that reduced task cancellation does not guarantee reduced overall execution time.
\end{abstract}


\section{Introduction}

%
%The background for this runtime is ongoing research within the AllScale project \cite{}, where we study recursive nested parallelism for distributed systems.
It is widely accepted that compute clusters and supercomputers are transitioning towards systems of millions of compute units to satisfy the requirements of compute-intensive parallel scientific applications.
With this increase in compute components, a proportional decrease in the Mean-Time-Between-Failure (MTBF) across parallel executions will follow \cite{Schroeder2010,Zheng2012}, which would make highly scalable parallel application runs infeasible without integrating resilience.
A new design of runtimes and resilience protocols is required to address the issue of recovering from failures.

In this manuscript, we design a resilience protocol for recovery from node failures. Such failures are among the most challenging ones to target, since they can not be detected by the failing node.
There are two major classes of recovery techniques from node failures, each of them employing a different degree of redundancy. In the presence of sufficient hardware resources, replication (which requires time/space redundancy) could be used so that failures could be masked altogether, often without any runtime overhead.
Alternatively, a variety of checkpoint/restart (C/R) strategies can be used \cite{Elnozahy2002,Egwutuoha2013}; they introduce time redundancy due to the rollback of execution but require fewer additional resources.
Our recovery strategy fits into the C/R recovery strategies.
However, unlike most of the related work in high-performance computing, we do not design a resilience strategy on top of message-passing libraries.
Instead, we build on top of dynamic task-based runtimes.
Our requirements on the runtime are as follows:
\begin{itemize}
\item Dynamic load balancing strategies, such as task stealing, are preferred
\item Task dependencies across a distributed execution are supported
\item Global data is migrated transparently by the runtime
\item The runtime supports creating partial executions which work on data replicas without affecting global data
\end{itemize}

As we shall explain, the last requirement is essential if we wish to enable dependency-aware rollback instead of rollbacks to the last checkpoint.

The goal of this work is to design a resilience protocol for checkpoint/restart strategies; ideally, the protocol does show some advantages over existing MPI solutions.
%It can be applied to resilient extensions of languages, including resilient Chapel variants \cite{Panagiotopoulou2016} or Resilient X10 \cite{Cunningham2014}.
%The developer works explicitly only with data, tasks, and task dependencies.
Unique to our work, we use task logging, which allows us to reduce the rollback of tasks at each worker during recovery.
This scheme can only be applied to task-based distributed systems, where task dependencies are explicitly defined.
We see task logging as the counterpart of message logging for message-passing protocols.
%The latter are very complex to design and implement, and not needed in task-based runtimes due to the explicit nature of task dependencies.

Our use case is an existing recursively decomposed stencil algorithm \cite{Frigo2005}.
We extend the stencil algorithm to distributed executions, while retaining its recursive decomposition scheme.
%We do this with the reasonable expectation that the processing of stencil fragments in different cache segments in shared memory translates into processing of even larger stencil fragments on different nodes, with similar efficiency.
Once a large enough stencil fragment is allocated to a node, further decomposition in shared memory can follow, as done very efficiently by popular implementations like Pochoir \cite{Tang2011}.

%The data pattern of checkpointing cannot be generalized, since it differs for each application kernel arbitrarily.
%The implicitly coordinated checkpointing relies on a pre-execution analysis about the conditions for a consistent checkpoint for stencil codes, which is tranferred into local and globally uncoordinated checkpointing.
%The source/target scheme of checkpointing can be generalised.  
We employ an in-memory neighbor checkpointing scheme we call guard/protectee scheme (similar e.g. to the buddy scheme of Charm++) in our prototype; it is generic and can be used for any application checkpointing scheme.

The novel contributions of this work are:
\begin{itemize}
\item A task logging protocol in addition to data checkpointing
\item A dependency-aware rollback, based on the task logs and task dependencies provided by any kernel
\item A simulator of distributed task-based runtimes for the validation and evaluation of resilience protocols.
\item A checkpoint/restart mechanism for distributed execution of stencil codes, influenced by the recursive formulation of stencils in related work
\end{itemize}
%The task logging and dependency-aware rollback are generally applicable to any application kernels expressed as tasks for task-based runtimes.
%The C/R protocol is applicable only to the underlying use case, stencil kernels.
The simulator of the proposed scheme demonstrates reduction in the cancelled tasks for runs with a handful of node failures on hundreds of nodes.
This reduction leads to 1.3--18\% less aggregated task processing time, and 1--10\% faster overall execution time, depending on the frequency of checkpointing.
%
%Despite the fact workers that the proposed checkpointing is uncoordinated, the workers individually checkpoint at the same timesteps in execution (without coordinating with each other).
%This does pose similarity to coordinated checkpointing strategies.
%For all such strategies, Daly/Young formula is well established.
%One central challenge we face is Young/Daly formulation for coordinated checkpointing, which has been validated both in theorety and practice, assumes constant checkpoint duration.
%This assumption does not hold in runtimes which allow arbitrary scheduling of tasks.

%There are two levels in our resilience manager:
%\begin{itemize}
%\item The general aspects include the guard/protectee scheme between workers, as well as checkpointing of tasks, which enables unique reduced rollback enabled by analyzing task dependendencies.
%\item The kernel-centric aspects include the checkpointing of data associated to tasks; this includes the prerequisite to using the efficient implicitly coordinated checkpointing, which we may only do having analyzed what builds a consistent checkpoint.
%\end{itemize}

%The recursive and hierarchical nature of our approach has various implications -- in our work we can use recursion levels to define the checkpoint frequency. In their work, each domain crossing needs to be checkpointed.
%Another work with the similar goal of transparent fault tolerance is X10-FT \cite{Hao2014}.
%Their solution differs in various ways, including the lack of control over checkpointing frequency.

The paper is organised as follows: In Sect. \ref{sec:related-work}, we present the related work.
In Sect. \ref{sec:guard-protectee}, we overview the guard-protectee scheme, followed by the checkpoint and task logging in Sect. \ref{sec:cp-protocol}.
We introduce the dependency-aware protocol in Sect. \ref{sec:recovery}.
We then introduce a stencil code (Sect. \ref{sec:kernel}), which serves as a foundation for illustrating the resilience protocol (Sect. \ref{sec:stencil-recovery}).
We continue by presenting a discrete-event simulator in Sect. \ref{sec:simulator}, and the evaluation it delivers in Sect. \ref{sec:experiments}.
We then discuss some issues in Sect. \ref{sec:discussions}, and conclude the paper in Sect. \ref{sec:conclusion}.

\section{Related Work}
\label{sec:related-work}
In this section, we first describe the related work in the MPI domain, followed by related work in task-based runtimes.
We also list the stencil code used as a case study.
We add that our work does not consider downtime, recovery time, or failure during recovery; these have been considered in related work such as \cite{Daly2006,Bosilca2014unified}.

\subsection{MPI Domain}
Gamell et. al. \cite{Gamell2014} integrates fault tolerance into a highly scalable MPI-based stencil code implementation called S3D.
Their checkpointing scheme uses implicitly coordinated checkpointing, which shows excellent scalability by avoiding global coordination.
The background for their checkpointing scheme is the simple and efficient idea of using in-memory checkpointing on peer nodes \cite{Zheng2012}.
The rollback in the S3D code is a classic rollback to the last globally consistent checkpoint.
In comparison, we also use in-memory checkpointing, and our checkpointing is also implicitly coordinated.
Our checkpoints use a different partitioning scheme; also, in contrast to any MPI solution, we employ task logging, which allows reduced rollback, based on the explicitly defined task dependencies in task-based runtimes.
In the MPI domain, reducing rollback is also possible -- through uncoordinated checkpointing and message logging.
As summarised in a seminal survey paper \cite{Elnozahy2002}, in message passing ``log-based rollback recovery in general enables a system to recover beyond the most recent set of consistent checkpoints''. 
This is a very challenging task, which often produces the ``domino effect'' resulting in restarting the entire application.
A recent MPI library extension \cite{Guermouche2011} integrates message logging with uncoordinated checkpointing to achieve efficient rollback; as a result, it reduces the number of restarting processes compared to coordinated checkpointing by a factor of 2.
We should note that this contribution is rather complex to implement, and applies to a particular class of MPI applications, called send-deterministic applications.

\subsection{Task-Based Runtimes}
The short-lived distributed version of Cilk \cite{Blumofe1995}, called Cilk-NOW \cite{Blumofe1997}, is an early example of adding fault tolerance for node failures in task-based runtimes.
The protocol is based on a centralised master-slave model, and so it differs significantly from the proposed decentralised protocol of this work.
More recently, efforts have been made to integrate resilience into task-based languages such as X10 \cite{Hao2014,Cunningham2014}, Chapel \cite{Panagiotopoulou2016}, and Charm++ \cite{Zheng2004,Huang2004}.
By relying on checkpointing, Charm++-FTC and X10-FT are closer related to our work than Chapel, which proposes replication instead.
In addition, the Charm++-FTC ``buddy'' scheme is similar to our guard/protectee scheme.
Across all novel schemes, including ours, in-memory checkpointing is assumed.

Leaving checkpointing techniques aside, our recovery scheme is new; we are not aware of existing work in the HPC domain for task-based runtimes employing task dependencies to reduce rollback in case of node failures.
%As we outlined in the introduction, it is our intention to implement this resilience protocol within HPX \cite{Kaiser2016}.

In the domain of task-based graph processing, a similar idea has recently been proposed \cite{Xu2016}; the authors implement an efficient ``confined recovery'' strategy by combining uncoordinated checkpointing and dependency logs.
Their results show significant improvement over global rollback.

\subsection{Stencil Kernel}
Our use case in prototyping the resilience protocol is a stencil kernel, which is the foundation of various scientific applications.
Importantly, we are influenced by the elegant recursive formulation of a stencil kernel~\cite{Frigo2005}, which has been efficiently implemented in shared memory in the Pochoir stencil compiler \cite{Tang2011}.

\section{Guard/Protectee Scheme}
\label{sec:guard-protectee}
We first present a guard/protectee scheme for all worker nodes, which is generally applicable and essential for the implemented recovery strategies.
The background for introducing a guard/protectee scheme is the simple and efficient idea of using in-memory checkpointing on peer nodes \cite{Zheng2012}.
%This checkpointing scheme was subsequently implemented efficiently for stencil codes \cite{Gamell2014}.

The guard/protectee scheme is generally applicable for checkpointing schemes, and not limited to stencil codes.
It defines:
\begin{itemize}
\item the source and target node of a checkpoint
\item the source and target of detecting faults, and the responsible node for recovery
\end{itemize}
In this work we do not deal with detection, however we remark that the guard/protectee scheme can conveniently be combined with a distributed heartbeat protocol implementation.
We use the guard/protectee scheme only for checkpointing and recovery in this work.
Each worker has a protectee, and a guard; they may or may not be the same, depending on the chosen strategy.
For example, in our prototype we employ a ring scheme: All workers have their guard on their right, and their protectee on their left (the first and last worker also being ``connected'').
Regardless of the chosen scheme, it always holds for any 2 workers $A$ and $B$, \[ guard(A) = B \Leftrightarrow protectee(B) = A\]
In this case, $A$ always sends its checkpoints and task logs (we detail them later) on node $B$.

Logging and checkpointing are  blocking on the sender node; a save operation into remote memory needs to complete before continued execution.
These operations do not block task processing at the guard node.
Therefore, no global coordination is required.
The in-memory checkpointing scheme scales to many thousand cores, as demonstrated for a supercomputer setting with an MPI application~\cite{Gamell2014}.
Still, in case of memory limitations, persistent storage on disk could be used instead of a guard/protectee scheme.

\section{Task Logging and Data Checkpoints}
\label{sec:cp-protocol}
In this section we present an uncoordinated checkpointing protocol, which stores at its guard:
\begin{itemize}
\item task logs (application independent)
\item data checkpoints (application dependent)
\end{itemize}
When storing a backup of task closures (we borrow the term from early Cilk prototypes \cite{Blumofe1995}), we use the term task logging.
When storing a backup of application-specific data, which is associated to a task, we use the term data checkpoints.
Our scheme for task logging and data checkpointing is outlined in Alg. \ref{alg1}:
\begin{algorithm}
\begin{algorithmic}
\State $closure \gets \Call {Get\_Closure}{t}$
\State $data \gets \Call {Get\_Data}{t}$
\If {$ granularity(t) \geq T_L $}
\State \Call{send\_task\_log}{\textit{closure}}
\EndIf
\If {$t$ entry task at $T_C$ boundary}
\State \Call{send\_checkpoint}{\textit{data}}
%\State \Call{swap\_buffers}{}
\EndIf
\State \Call{process}{\textit{T}}

\end{algorithmic}
\caption{Task logging and data checkpointing.}
\label{alg1}
\end{algorithm}

We do not log or checkpoint tasks at fine-grained levels to reduce overhead.
Instead, we only log tasks down to level $T_L$,  the minimal task level at which a worker may steal a task from another node. This task may be further decomposed locally, spawning child tasks for shared-memory processing.
For example, this shared-memory fine-grained decomposition may be implemented via Cilk Plus as in our use case \cite{Tang2011}.
The granularity of $T_L$ is the responsibility of a distributed scheduler; our resilience strategy uses it, but does not determine its value.

%\subsection{Checkpointing Protocol}
Each worker first fetches the data (\verb1Get_Data1) and closure (\verb1Get_Closure1) of a $T_L$-level task it steals.
This is either a local memory copy, or a network transfer. 
Task logging is generic, and can be applied to any application kernel implemented via tasks; as we will detail, task logging is required for our dependency-aware rollback, which improves resource utilisation.

Application data checkpoints are needed to support any C/R technique.
We denote the chosen checkpoint level with $T_C$, and we experiment with various such levels.
The finest grained checkpoint level is 1, and it means $T_C = T_L$ (we assume $T_C \geq T_L$).
Thus, a worker only conditionally checkpoints data, exactly at the entry level of a $T_C$-level task.
We will detail $T_C$-level checkpointing on the example of stencils in Sect. \ref{sec:stencil-rec}.

%We assume that each worker may steal tasks of exactly level $T_L$, 

%Based on the excellent scalability it shows for stencil codes \cite{Gamell2014}, and in agreement with our proposed guard/protectee scheme between nodes, the main memory of the guard node is the preferred destination of a checkpoint.

\section{Dependency-Aware Recovery}
\label{sec:recovery}

We assume here single-node failure at a time; an extension to multiple node failures is simple, but more costly, and consists of extending the guard-protectee scheme to a multiple guards-multiple protectees scheme.
In case of a node failure, in contrast to the uncoordinated checkpoint strategy, the recovery strategy is coordinated and blocking across all workers.
%For illustration purposes, consider Fig. \ref{fig:consistent-cps}.
%Let the failure of W2 be detected.
%Let also the checkpointing granularity be $T_C = 2$, which consists of 4 $T_L$ level triangles.
%A $T_L$-level task in red does not complete due to the node failure.

The recovery begins with the guard node fixing the guard/protectee scheme globally.
This operation does not require all workers to participate -- it updates only the guard, the failed worker's protectee, and optionally the replacement worker (if one is available).
Then, one of two recovery strategies are possible, which we describe and compare accordingly: a standard rollback to the last globally consistent checkpoint, or the proposed dependency-aware rollback.
%\item All surviving workers coordinate to establish which local checkpoints build a consistent $T_C$-level checkpoint. This is not possible during checkpointing. For example, consider that a node may checkpoint and process both T3 and T4 in Fig. \ref{fig:double-buffer}. It has no notion which consistent checkpoint has been completed -- checkpoint 1 or checkpoint 2.
\subsubsection{Standard rollback} 
On each worker cancel all tasks past the last consistent checkpoint and reschedule them. The failed node's guard also keeps a task set $L_1$ of all $T_L$-level tasks of the failed protectee, which were started and not backed up.
The guard reschedules all of them.
\subsubsection{Dependency-aware rollback}
\begin{itemize}
\item The guard node keeps the described task set $L_1$. From $L_1$, a task set $L_2$ is created, consisting only of the local checkpoint baseline that will produce $L_1$.
We compute the task set $L_3$ as exactly the tasks leading up to the $L_1$ tasks from the checkpoint baseline $L_2$. We enclose the algorithm in Alg. \ref{alg2}.
\item  All $L_3$ tasks are rescheduled in order to reproduce the failed task again.
The distributed runtime is \textbf{strictly required} to run all $L_3$ tasks on distinct data replicas without affecting the global data.
This may be done via special annotation of the associated data to restarted tasks as independent and short-lived global data, which exists until all $L_1$ tasks are restarted.
%\item Every surviving worker $w_i$ creates a queue $L_3$, which consists of all tasks with dependencies on $L_2$. These tasks are illustrated in orange in Fig. \ref{fig:consistent-cps}.
%They form stage 2 of cancellation.
%The stage 2 cancelled tasks are dependent tasks on any of the red or orange tasks.
%This means that the coherent global data is not guaranteed anymore if only stage 1 cancellation happens without stage 2 cancellation.
%We do not further execute this direction in this work, but we believe it is likely for a runtime supporting multiple and incoherent copies of global data (and some coherence protocol to fix this) to avoid stage 2 cancellation altogether.
\end{itemize}
%(Note: A cancel operation on a task either returns immediately (already complete task), or blocks until the started task completes). All cancelled tasks are rescheduled by their worker.

\begin{algorithm}
\begin{algorithmic}
\State $L_3 \gets \emptyset$ \Comment $L_3$ - tasks to restart
\For {$ t_1 \in L_1$} \Comment $L_1$ - failed tasks
\For {$ t_2 \in L_2$} \Comment $L_2$ - checkpoint baseline
\For {$ t_3: \forall$ tasks}
\If {$t_1$ depends on $t_3$ and $t_3$ depends on $t_2$}
\State $L_3 \gets L_3 \cup t_3$
\EndIf
\EndFor
\EndFor
\EndFor
\State \Return $L_3$
\end{algorithmic}
\caption{$L_3$: Set of cancelled tasks to recover failed tasks $L_1$}
\label{alg2}
\end{algorithm}
We do not claim or prove in this work that Alg. \ref{alg2} minimises the number of cancelled tasks.
However, the underlying idea is exactly that -- to minimise task cancellation across all nodes, based on task dependencies.
As we demonstrate in the experimental section on a use case, the dependency-aware rollback meets our expectations; less tasks need to be restarted than in the case of standard rollback to a consistent checkpoint.
This immediately translates into better utilisation of compute resources.
In general, reduced rollback does not necessarily transfer into faster overall execution time,
However, it consistently resulted in faster execution time in our evaluation.

\section{Recursive Decomposition of Stencil Kernel}
\label{sec:kernel}
\subsection{Formulation}
\begin{figure}[!htb]
  \centering
  \includegraphics[width=\linewidth]{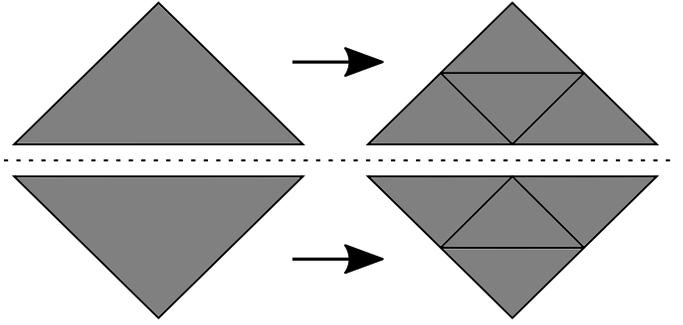}
  \caption{Decomposition example of tasks into smaller ones}
  \label{fig:decomposition}
\end{figure}

A powerful recursive formulation of the stencil has been proposed in earlier work \cite{Frigo2005}.
Our proposed decomposition is a variation on this work.
Similar to the original work, we implement each further decomposition as a recursion step in space (space cut) or time dimension (time cut).
Due to space constraints, we do not outline the entire algorithm.
We restrict this work to one-dimensional stencils.
We assume the initial stencil is of rectangular shape (\verb_<space,time>_ dimension input), and gets decomposed with space and time cuts into regular triangular shapes.
Once the initial rectangular shape is decomposed into these shapes, a combined space/time cut can happen as illustrated in Fig. \ref{fig:decomposition}, decomposing in each step a task into 4 sub-tasks.
Once the task granularity is small enough -- task level $T_L$ -- any further decomposition or processing is subject to the shared-memory implementation within a node.
On the example of the Pochoir stencil compiler \cite{Tang2011}, Cilk Plus provides such an implementation.

%The recursive decomposition, down to node level $T_L$, is outlined in Alg. \ref{algo:stencil}.
%\begin{algorithm}
%\begin{algorithmic}
%\Function{walk\_rectangle}{$x_0,x_1,t_0,t_1$}
%\State $\Delta{T}$ = \Call{Max}{$(x_1-x_0),(t_1-t_1)$}
%\If {$\Delta{T} \leq T_L$}
%\State split into 2 triangles
%\State \Call{process\_triangles}{\dots}
%\Else
%%\If{$x_1 - x_0 \geq \Delta{T} $}
%%\State \Call{walk}{$x_0,\frac{x_1-x_0}{2},t_0,t_1$}
%%\State \Call{walk}{$\frac{x_1-x_0}{2},x_1,t_0,t_1$}
%%\Else
%%\State \Call{walk}{$x_0,x_1,t_0,\frac{t_1-t_0}{2}$}
%%\State \Call{walk}{$x_0,x_1,\frac{t_1-t_0}{2},t_1$}
%\If {$ x_1 - x_0 > t_1 - t_0 $}
%\State space cut into two 
%\State call \Call{Walk\_rectangle}{$\dots$} on each cut
%\Else
%\State time cut into two
%\State call \Call{Walk\_rectangle}{$\dots$} on each cut
%\EndIf
%\EndIf
%\EndFunction
%\end{algorithmic}
%\caption{Outline of recursive stencil algorithm as a 1D triangular variation on related work \cite{Frigo2005}.}
%\label{algo:stencil}
%\end{algorithm}

\subsection{Task Dependencies}
\begin{figure}[!htb]
  \centering
  \includegraphics[width=\linewidth]{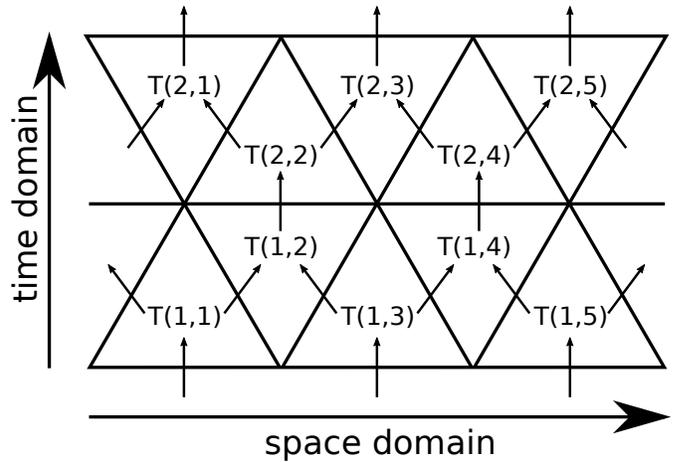}
  \caption{Task dependencies for 1D stencils}
  \label{fig:task-dep}
\end{figure}

The task dependencies for task stencils are visualised for a trivial 1D stencil in Fig. \ref{fig:task-dep}.
The dependencies can be formalised as follows:
\begin{itemize}
\item Each task of $\bigtriangleup$ shape, e.g. T(1,1), depends on exactly one element:
$T(t,n) \leftarrow T(t-1,n)$
%\kostas{$T(t,x_0,x_1) \leftarrow T(t-1,x_0,x_1)$}
\item Each task of $\bigtriangledown$ shape, e.g. T(1,2), depends on exactly two elements:
$T(t,n) \leftarrow T(t,n-1)  $ and $T(t,n) \leftarrow T(t,n+1)$
%\kostas{ $T(t,x_0,x_1) \leftarrow T(t,x_0-dx,x_1-dx)  $ and $T(t,x_0,x_1) \leftarrow T(t,x_0+dx,x_1+dx)$ }
%\item The border elements sligtly differ from the above. A border task of regular orientation passes on one less dependency on (see T1). A border task of upside-down orientation receives one less dependency (see T4).
\end{itemize}

\subsection{Task Closure and Task Data}
A task closure contains the information needed for the runtime to schedule task execution.
It may contain:
\begin{itemize}
\item global task identifier
\item task dependencies
\item task input arguments
\end{itemize}

For example, consider task T(1,1) which might have been created by the evaluation of a \verb1process_task1 call in the space range \verb_(1,1)_ and time range \verb_(1,1)_(Fig. \ref{fig:task-dep}).

Its global identifier may be \verb2<ID-1>2, its input arguments \verb_(1,1,1,1)_.
Each task is always associated with a fragment of global data (here: the stencil).
The global data associated to T(1,1) is not contained in the closure.
The body of T(1,1) will update a region of the stencil.
The location of this range is managed by the runtime, which fetches it (if needed) from another worker.
Message passing takes place transparently to the task body (still, we model it as a network transfer in our evaluation).
If task T(1,2) is executed at another worker, the runtime will transparently fetch the global data associated with it.

%\section{Iterative Map/Reduce Kernel}
%\subsection{Formulation}
%An iterative Map/Reduce based runtimes has been proposed for example in \cite{Ekanayake2010}.
%This covers a class of algorithms, among others including e.g. implementations for k-means, Pagerank etc.

%One possible formulation of the iterative map/reduce algorithm is:

%\begin{algorithmic}
%\For {$i = 1 \dots N$}
%\State Break down problem space into $M$ $T_L$ sized %tasks
%\State \Call{Map/Reduce}{$\dots$}
%\EndFor
%\end{algorithmic}
%\subsection{Task Dependencies}
%The task dependencies for iterative map/reduce methods are shown in Fig. \ref{fig:task-dep2}.
%As shown in the figure, we recognize that often map and reduce operations can be parallelized with no dependencies (for example: T1,T2,T3 being all tasks for map workloads in one iteration).
%However, each iteration step needs to complete before the next one can (hence the many task dependencies).
%This is enforced by the fact that combining results in map/reduce is an implicit global operation.
%\input{task-deps2}

\section{Resilience Protocol for Stencil Codes}
\label{sec:stencil-recovery}
\subsection{Stencil-Specific Data Checkpoints}
\label{sec:stencil-rec}
\begin{figure}[!htb]
  \centering
  \includegraphics[width=\linewidth]{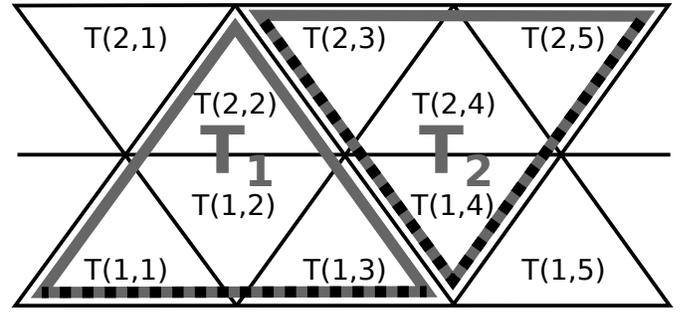}
  \caption{Illustration of checkpointing boundaries at task entry (dashed lines) at checkpoint level 2.}
  \label{fig:double-buffer}
\end{figure}

In this section, we outline a novel checkpointing scheme for stencil codes, which differs from existing checkpointing schemes based on iterative methods.
Existing schemes rely on periodic checkpointing in time.
The proposed scheme in our work is the recursive decomposition of a stencils for shared memory~\cite{Frigo2005}.
However, since it simply partitions the task set into checkpoint tasks and regular tasks, it can probably be applied to iterative kernels as well.
There is a strong motivation for our scheme: it allows for a very fine granularity, which is increasingly important in C/R techniques, due to the lower cost of in-memory checkpointing.
The finest-grained checkpoint level is $T_L$ -- the level of task logging.
The next level, checkpoint level 2, includes 4 $T_L$ triangles,
and is illustrated in Fig. \ref{fig:double-buffer}.
Visually, the checkpoints are the ``lower'' entry lines of the $T_C$ triangles (since computation progresses vertically from bottom to top).
The upper limit to the granularity of data checkpoints is the time or space dimension.
For checkpoint levels 1--6 and a 256x256 time/space grid, we list $T_C$ and $T_L$ characteristics in Table \ref{tab:cp-gran}.
We remark that the data checkpoint technique for the stencil kernel cannot be generalised due to the application-specific nature of globally consistent data checkpoints.
There is an important difference to iterative approaches in the proposed stencil checkpointing scheme:
\begin{itemize}
\item Existing iterative checkpointing schemes use the global time line for periodic checkpointing
\item Our scheme uses $T_C$-level triangles as boundaries of local checkpoints
\end{itemize}

\begin{table}
\begin{tabular}{|c|cccccc|}
\hline
Checkpoint Level & 1 & 2 & 3 & 4 & 5 & 6 \\
\hline
\begin{varwidth}{2.5cm}$T_L$-level triangles in a $T_C$ triangle\end{varwidth} & 1 & 4 & 16 & 64 & 256 & 1024 \\
\hline
    \begin{varwidth}{2.5cm}Local checkpoint count ($T_C$ triangles) for $256^2$ $T_L$-level tasks \end{varwidth}& 65536 & 16384 & 4096 & 1024 & 256 & 64 \\
\hline
\end{tabular}
\caption{Checkpoint levels and $T_C$/$T_L$ triangles}
\label{tab:cp-gran}
\end{table}

The illustrated segment in Fig. \ref{fig:double-buffer} has 2 $T_C$-level checkpoints: $T_1$'s entry boundary, which includes the global data associated to $T(1,1)$ and $T(1,3)$; $T_2$'s entry boundary, which includes the global data associated to $T(2,3)$, $T(1,4)$ and $T(2,5)$.
Of course, larger granularity means less tasks need to be checkpointed in comparison to the total task count.

The checkpointing scheme is efficient at runtime, and efficient in its buffer requirements.
Due to the existing stencil dependencies, it is sufficient to use double buffering at each worker.
We store at most one complete and one incomplete checkpoint of a piece of global data at a time.

\subsection{Illustrating Dependency-Aware Recovery}
On the example of the stencil kernel, the dependency-aware rollback can be easily illustrated.
\begin{figure}
\centering
\begin{tikzpicture}[node distance=0.95cm]
\tikzstyle{block} = [%
   rectangle, draw,thick,
    text centered,font=\tiny,rounded corners]

%\node[block,blue](t1check) {CP};
\node[block,dashed](t1a) {$T(1,1)$};
\node[block,right of=t1a](t2) {$T(1,2)$};
\node[block,right of=t2](t5) {$T(2,2)$};
\node[block,dashed,right of=t5](t6a) {$T(2,3)$};

%\node[block,blue,below of=t1check](t2check) {CP};
\node[block,dashed,below of=t1a](t3a) {$T(1,3)$};
\node[block,dashed,right of=t3a](t4) {$T(1,4)$};
\node[block,right of=t4](t7) {$T(2,4)$};
\node[block,dashed,right of=t7](t8a) {$T(2,5)$};
\node[below of=t3a](tleft1){};
\node[below of=t8a](tright1){};
\draw[->] (tleft1.west) -- node[label={Timeline}]{} (tright1.east);
\begin{scope}[xshift=4cm]
\node[block,dashed](t1) {$T(1,1)$};
%\node[block,blue,right of=t1](t2check) {CP};

\node[block,dashed,right of=t1](t3) {$T(1,3)$};
\node[block,right of=t3](t2) {$T(1,2)$};
\node[block,dashed,below of=t2](t4) {$T(1,4)$};
\node[block,right of=t2](t5) {$T(2,2)$};
\node[block,right of=t4](t7) {$T(2,4)$};
\node[block,dashed,right of=t5](t6) {$T(2,3)$};
\node[block,dashed,right of=t7](t8) {$T(2,5)$};
\node[right of=tright1](tleft2){};
\node[below of=t8](tright2){};
\draw[->] (tleft2.west) -- node[label={Timeline}]{} (tright2.east);
\end{scope}
\node[dashed,block] (legend1) at ($(current bounding box.east)+(0,2.5)$) {CP task};
\node[block,below of=legend1,node distance=0.5cm] (legend2) {\begin{varwidth}{4cm}Regular task\end{varwidth}};
\begin{pgfonlayer}{background}
\draw[fill,gray!30] ($(t1a.west)+(-0.25,0.25)$) -- node[label={north:\color{black}worker1}]{} ($(t6a.east)+(0,0.25)$) -- ($(t6a.east)+(0,-0.25)$) -- ($(t1a.west)+(-0.25,-0.25)$);
\draw[fill,gray!30] ($(t3a.west)+(-0.25,0.25)$) -- node[label={north:\color{black}worker2}]{} ($(t8a.east)+(0.25,0.25)$) -- ($(t8a.east)+(0.25,-0.25)$) -- ($(t3a.west)+(-0.25,-0.25)$);
\draw[fill,gray!30] ($(t1.west)+(0,0.25)$) -- node[label={north:\color{black}worker1}]{} ($(t6.east)+(0.25,0.25)$) -- ($(t6.east)+(0.25,-0.25)$) -- ($(t1.west)+(0,-0.25)$);
\draw[fill,gray!30] ($(t4.west)+(-0.25,0.25)$) -- node[label={north:\color{black}worker2}]{} ($(t8.east)+(0.25,0.25)$) -- ($(t8.east)+(0.25,-0.25)$) -- ($(t4.west)+(-0.25,-0.25)$);
\end{pgfonlayer}
\end{tikzpicture}
\caption{Two different processing schedules via work-stealing (tasks as in Fig. \ref{fig:task-dep}). Each execution results not only in slightly different processing duration, but also different checkpointing duration.}
\label{fig:two-schedules}
\end{figure}
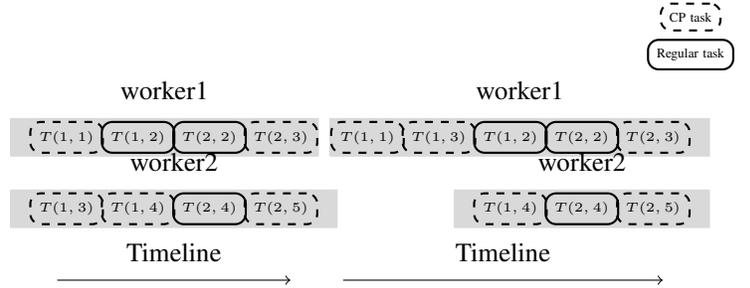

\begin{figure}[!htb]
  \centering
    \includegraphics[width=\linewidth]{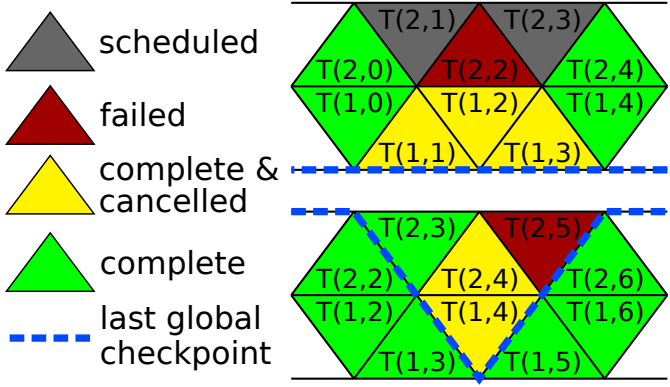}
  \caption{Illustration of reduced rollback for the 2 alternative checkpoint boundaries. Cancelled tasks (in yellow) can reproduce the failed tasks (in red). No other complete tasks (in green), or scheduled tasks (in grey) need to be cancelled.}
  \label{fig:consistent-cps}
\end{figure}

We play through the C/R protocol with dependency-aware rollback for any of the scheduling sequences shown in Fig. \ref{fig:two-schedules}, using 2 workers.
The failed, complete, and cancelled tasks, are illustrated accordingly in Fig. \ref{fig:consistent-cps} (below).
We assume the second worker fails during processing of task $T(2,5)$.
We use terminology for the dependency-aware rollback as in Alg. \ref{alg2}.

W2 crashes while processing $T(2,5)$. The failure is eventually detected and broadcast to all workers.
W1 decides a worker W3 from a pool may join, and fixes the guard/protectee scheme in sync with W3.
W1 then establishes that $L_1 = \{T(2,3),T(1,4),T(2,5)\}$.
W1 has task set $L_2 = \{T(1,4),T(2,4),T(2,5)\}$, all successfully initiated on W2.
It evaluates the dependencies and generates the list of tasks to cancel and restart as $L_3 = \{T(1,4),T(2,4),T(2,5)\}$ (cancelled tasks marked in yellow, failed task in red in Fig. \ref{fig:consistent-cps}).
No other tasks are restarted, since no other dependencies exist.
$T(2,3)$ on W1 never was initiated, so it needs no rescheduling.
All $L_3$ restart tasks are cancelled and recomputed from the consistent checkpoint, with an annotated distinct replica of the global data, until they satisfy the $T(2,5)$ dependencies.

%\subsubsection{Right}
%W2 crashes while processing $T(2,5)$ The failure is detected and broadcast.
%W1 decides a worker W3 from a pool may join, and fixes the guard/protectee scheme in synch with W3.
%W1 establishes that $L_1 = \{T(2,3),T(1,4),T(2,5)\}$.
%W1 has task set $L_2 = \{T(1,4),T(2,4),T(2,5)\}$ as initiated on W2.
%It cancels and reschedules all of them, since they depend on the latest checkpoint.
%\item W1 cancels and reschedules only $T(2,3)$, since it is the only task with dependency on the latest consistent checkpoint

%\input{consistent-cps}
The example illustrates clearly that a global rollback to a checkpoint is not needed, and this can be established through the task dependencies.
%Second, the nature of dynamic task-based runtimes is extremely complex, and even minor differences in scheduling tasks has implication on how the resilience protocol will recover a consistent state, and which tasks will be cancelled.
%Indeed, due to the dynamic scheduling policies, we may observe variable real time checkpoint duration.
%Only the underlying task dependencies of stencils restrict the potentially significant variation in the execution point at which a worker checkpoints a task at stencil time $T$.
%This has far reaching consequences to the estimate of checkpointing duration and checkpointing intervals, as we will detail in Sect. \ref{sec:young-daly}.

\section{Resilience Simulator RS}
\label{sec:simulator}
We have implemented in Python a discrete-event simulator called RS (The code is open sourced under \cite{rs-repo}).
A discrete-event simulator has some desirable properties which reduce the complexity of experiments.
For example, we control and can reproduce varying settings.
Also, the entire overhead in handling resilience (e.g. management of task queues or synchronisation of tasks) does not affect the simulated runtime.
This allows to observe the effect of failures and recovery in a more controlled environment.

Besides the technical reasons for using a discrete-event simulator, there are two main reasons behind this development:
\begin{itemize}
\item It is challenging to verify the correctness of a resilience protocol for task-based runtimes. Due to the potential dynamic load balancing of tasks, and the task dependencies, various scheduling policies are possible.
This becomes extremely complex once critical failures are introduced. A simulator can validate the resilience protocol is correctly implemented, and that the recovery results in the completion of all tasks.
\item It is difficult to provide any performance analysis of task-based runtimes in the presence of faults.
The simulator can provide experimental insights into performance-related questions, including in relation to optimal checkpointing (see Sect. \ref{sec:discussions})
\end{itemize}

We depend on SimPy \cite{SimPy}, a process-based discrete-event simulation framework based on standard Python and its generator functions.
Some of the main features of RS include:
\begin{itemize}
\item many-worker runs
\item parallel list processing as approximation for work stealing
\item support for various application kernels (including the 1D stencil kernel)
\item support for various scheduling policies
\item various checkpointing policies
\item failure and recovery scenarios
\end{itemize}

The main scenario we target in this work is the simulation of a distributed system of workers which can process tasks using work stealing, or similar dynamic approaches; the workers may experience node failures after which they need to recover via checkpoint/restart techniques.
In this section we summarise the implementation of the simulator.%, enlisting a simplified class diagram of its main classes in Fig. \ref{fig:simulator-uml-diagram}.

\subsection{Worker}
The worker class models the behaviour of each worker node. 
The main task of a worker is to steal and process tasks, and in addition to implement resilience policies.
One part of resilience is the consistent guard/protectee scheme.
Another part is the checkpoint/restart policy.

Each worker processes tasks from a global queue by simply calling generic methods like \textit{process}, which hide their kernel-specific logic in an implementation of the abstract \textit{WorkItem} class.
The Worker logic is entirely decoupled from the underlying task implementation.
A worker also periodically checkpoints tasks.

Failures are triggered as interrupts of SimPy processes, and can be graciously handled by each surviving worker in the exception handling block (which simply calls the function \textit{recover}).
The two implemented recovery methods are central to this work, and detailed in Sect. \ref{sec:recovery}.
%It includes task cancelling and task rescheduling.

\subsection{Queue and Scheduling Policies}
\label{sec:queue-and-scheduler}
Tasks are inserted into a global queue.
SimPy provides resources with a specified capacity.
We set the worker count to be the capacity; this enables all worker to process 1 task each, in parallel.
We use this as an approximation of work stealing.
We remark that in real work stealing approaches, each worker manages its own task queue; we oversimplify this concept, since the global queue abstracts away the entire synchronisation enabling work stealing between workers.

The scheduling policy is extremely important in any task-based distributed runtime.
We only provide one scheduling policy which is tailored to the stencil kernel.
We do this in order to guarantee an efficient processing of stencil $T_L$-level tasks, at the same time avoiding a potential deadlock.
All tasks are modelled as $T_L$-level tasks.
The stencil-specific scheduling policy is as follows:
\begin{algorithmic}
\For{$t$ in 0 \dots max-time}
	\State Append all $\bigtriangleup$ shape $T_L$-level tasks of time step $t$
	\State Append all $\bigtriangledown$ shape $T_L$-level tasks of time step $t$
\EndFor
\end{algorithmic}

On the example of Fig. \ref{fig:task-dep}, tasks \textit{T(1,1), T(1,3), T(1,5)} are added before \textit{T(1,2), T(1,4)}.
We were unable to devise a more efficient deadlock-free scheduling policy for stencils; a scheme employing a depth-first version was also tested, and delivered worse performance.

Independent of the scheme, the tasks are processed by workers in a FIFO order, each worker removing the first task from the queue once it has no work; it processes it only after the task dependencies have been resolved.

Importantly, each cancelled task is rescheduled into the global queue following the above logic to avoid deadlocks.
\subsection{WorkItem Implementations}
The WorkItem class is virtual, and each kernel needs to implement:
\begin{itemize}
\item the duration of tasks
\item task dependencies
\item the consistency of a checkpoint
\end{itemize}

All task dependencies are observed.
A task may only be processed if its dependency tasks are completed.
There are two possible implementations of this concept in SimPy -- signalling and busy waiting.
In this work, we use busy waiting at each worker to check when the dependent tasks are complete.

A checkpointing implementation cannot be generalised in our view; a checkpoint and its consistency always remains kernel-specific, and depends on how each kernel manages its data.
One of the most difficult implementation aspects was in fact the stencil-specific checkpoint policy, and the model of a consistent global checkpoint.
The main difficulty here is that the simulator is oblivious of any data.
Therefore, it requires tedious debugging and verification to ensure that the logic of local and global consistent checkpoints is correct.
In contrast to that, the task dependencies are easier to verify, and the simulator guarantees that they are always met.
This verification is very important, since apart from the global consistent checkpoint, we encountered two major issues when implementing a resilience protocol incorrectly: hanging execution or end of execution without successful task completion.
%For example, there is some similarity between these otherwise different kernels, and a horizontal cut through the data builds a consistent checkpoint.
%The implementation of 
%However, it is conceivable that other kernels are less tightly coupled in their data management, and may not even require a consistent checkpoint altogether.

The task processing, as well as data transfers or checkpoints in the simulator, are all timeouts of specified duration, as specified in the configuration file.

\subsection{Configuration File}
The simulator reads a configuration file as follows:

\begin{small}
\begin{verbatim}
[GENERAL]
Kernel= Stencil
WorkerCount = <WC>
CheckpointLevel = <CP>
Checkpoint = Y|N
Recovery = Default|Dependency
Fail = Y|N
MTBF = <MTBF>
Seed = <Random-Seed>
Scheduler= Horizontal
[Stencil]
BackupCost = 0.0013
ProcessCost = 7.1
StencilSize =128
Timesteps = 128
\end{verbatim}
\end{small}

The Kernel option specifies which kernel should be used, reading its corresponding configuration block.
Sect. \ref{sec:experiments} explains how we derive the values for processing a task (ProcessCost), and for transferring/checkpointing a task (BackupCost).
A non-uniform task duration is possible, but not implemented in this work.
The checkpoint level is detailed in Sect. \ref{sec:stencil-recovery}.
The recovery methods can be default recovery (from latest checkpoint), or our proposed dependency-aware recovery.
The MTBF option is the Mean-Time-Between-Failures, which we implement as a pseudo-random (using a specified random seed) generation of worker failures during execution, based on the Poisson distribution.
The only currently supported scheduling policy is explained in Sect. \ref{sec:queue-and-scheduler}.

\section{Evaluation}
\label{sec:experiments}
We detail what reference values we use for the simulation experiments.
All of these are derived from shared memory or cluster settings:
\begin{itemize}
\item We perform Pochoir benchmarks on a state-of-the-art server to set the duration of each $T_L$-level task. We use Pochoir since it provides various stencil codes of different dimensions, and since our recursive decomposition is a natural extension of this approach for distributed systems. We assume that the stencil code benchmarks (their runtimes are always in the order of few seconds) can be used as a reference for $T_L$-level tasks.
\item We estimate the communication cost of stencil data, and use it to model both data exchange and data checkpointing. Results from S3D \cite{Gamell2014} demonstrate that the network transfer is the main cost in the checkpointing phase.
We use a point-to-point throughput of 10 Gbps (representative for point-to-point communication on 10 GigE or Infiniband networks) as reference value.
We do not assume congestion in this work, since \textbf{(a)} we use uncoordinated checkpointing, and \textbf{(b)} various modern interconnects avoid congestion for multiple point-to-point channels.
\end{itemize}

%\subsection{1D Stencil}
%\subsubsection{Processing Duration}
%In Pochoir the APOP 1D stencil with 2 mill. cells and 10 thousand timesteps takes 4 seconds to compute on a 12-core (older) 2.6 GHz Xeon machine.
%We assume one such computation to be an element of level $T_L$.
%\subsubsection{Checkpoint Duration}
%We can approximate the checkpoint duration of $T_L$ tasks, which corresponds to 16 MB (2 mill doubles).
%We calculate it as approximately 1.28 milliseconds per $T_L$ task.

\subsection{1D Stencil via Pochoir}
\subsubsection{Processing Duration}
We benchmark one of the Pochoir kernels, a 1D stencil, \verb$heat_1D_NP$, with dimensions 200K cells and 200K time steps;
we assume one such computation to represent an element of level $T_L$.
We use a single-socket 16-core Xeon server, with a E5-2683 v4 processor and 2.1 GHz frequency, 40 MB LLC, and 256 GB RAM.
It takes $\approx{7.1}$ seconds to run the 200K x 200K kernel on all cores, with $\approx{5.18}$ time speedup compared to single-core execution.

\subsubsection{Exchange/Checkpoint Duration}
We approximate the communication and checkpoint cost by first computing the data size to checkpoint as $200K$ elements of type double, or 1.6MB.
For 10 Gbps throughput, this results in 0.00128s seconds per exchange/checkpoint (excluding latency).
To include the memory copy overhead at the remote node (which is an order of magnitude quicker), and the latency, we round up the estimate to 0.0013s (1.3 milliseconds).
%If this cost seems low, we refer to the S3D estimates, which display an even faster aggregate checkpointing speed on a supercomputer.

\subsection{Strong Scaling}
\begin{figure}
	\centering
	\includegraphics[width=0.4\textwidth]{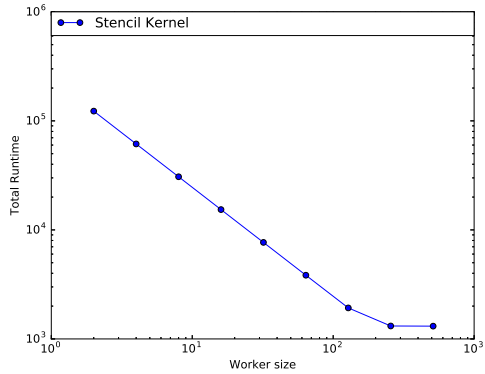}
	\caption{Strong scaling of the stencil kernel for a 128 cells x 128 time steps 1D stencil.}
	\label{fig:strong-scaling}
	\end{figure}

We first perform a few experiments to demonstrate the parallelization of task processing in the simulator.
We simulate a 128 array x 128 time step grid, or $128^2 T_L$-level task execution.
We arbitrarily set a 5 second duration per $T_L$-level task; the outlined scheduling policy is used, and checkpointing or failures are disabled.
Fig. \ref{fig:strong-scaling} shows the strong scaling capabilities of the simulator for a stencil kernel.
The experiments confirm the expectations.
Up to 128 workers, we observe ideal speedup.
For further increase, parallelization decreases and stalls due to the task dependencies, which force workers to wait.
We also measure fairness between workers, which is ideal, each worker processing the exact same number of tasks.
This is not surprising, since SimPy provides a semaphore which gives each worker a fair share of work.
	
\subsection{Comparison Between Default Rollback and Dependency-Aware Rollback}
	
The simulator allows us to compare various aspects of the major two rollback strategies described in Sect. \ref{sec:recovery}: the default rollback to the last consistent checkpoint, and the dependency-aware rollback.

We show results for $256^2$ $T_L$-level tasks, i.e. 256 $T_L$-sized stencils, and 256 time steps. We vary the checkpoint level $T_C$.
We use 128 worker nodes.
The used duration of a task and network transfer cost (equal to the checkpoint duration) are 7.1 seconds, and 1.3 milliseconds.

We set the MTBF to 30 minutes, using Poisson distribution to generate faults throughout the execution.
We use the same random seed across all experiments so that we compare runs under identical and reproducible failure scenarios.
We opt for the immediate replacement of a failed node by a new node.
We do not model the delay in detecting a failure, and in introducing a new node.
The recovery via rollback introduces recomputing of tasks, which is reflected in our evaluation.

\begin{figure}
\centering
\begin{minipage}{1.\textwidth}
\includegraphics[width=0.4\textwidth]{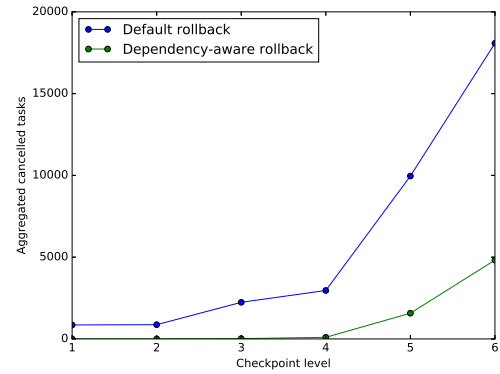}
\end{minipage}
\begin{minipage}{1.\textwidth}
\includegraphics[width=0.4\textwidth]{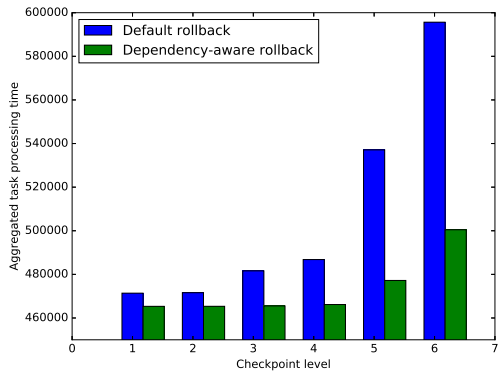}
\end{minipage}
\caption{Comparison between default and dependency-aware rollback. Above: Aggregated cancelled tasks. Below: Aggregated processing time (incl. recomputing).}
\label{fig:processing-time}
\end{figure}

For these settings, a run takes between 3600 and 4800 seconds to process all $256^2$ $T_L$-level tasks.
4 or 5 faults (an extra fault is generated for longer runs) are generated at random but reproducible times.
Fig. \ref{fig:processing-time} (top) shows the cancelled tasks for both strategies, Fig. \ref{fig:processing-time}(below) shows the aggregate processing time, for varying checkpoint levels.
When a fine-grained checkpointing is used ($T_C = 1$), the advantage of dependency-aware rollback is marginal, with 1.3\% less aggregated task processing.
This is intuitively clear: if we checkpoint often, we perform a minimal rollback even for the default strategy, and we reduce the margin to the intelligent recovery via dependency-aware rollback.
When using coarser checkpoint granularity, we gain in the dependency-aware rollback, since we cancel fewer tasks within a $T_C$-level triangle; we only cancel the tasks needed to reproduce the failed tasks.
For checkpoint level 6, the most infrequent checkpointing level, the aggregated procesing time is reduced by 16\% with the dependency-aware rollback.
Fig. \ref{fig:processing-time} (below) shows how this difference in cancelled tasks directly translates in reduced processing time (aggregated across all workers).
%\begin{figure}
%\begin{minipage}{1.\textwidth}
%\includegraphics[width=0.5\textwidth]{../current/1D/recoveries1}
%\end{minipage}
%\begin{minipage}{1.\textwidth}
%\includegraphics[width=0.5\textwidth]{../current/1D/recoveries2}
%\end{minipage}
%\caption{Simulation of complete rollback and proposed rollback based on task dependency information. Left: Total cancelled tasks. Right: Total runtime (secs)}
%\end{figure}

\subsection{Overall Runtime and Profiling}
\label{sec:overall-runtime}
\begin{figure}
\centering
\includegraphics[width=0.5\textwidth]{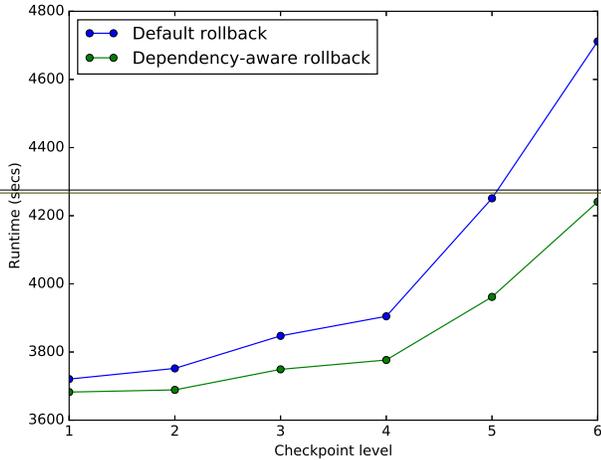}
\caption{Overall execution time: Reduced rollback translates in reduced overall execution time in our evaluation.}
\label{fig:runtimes}
\end{figure}

Reducing rollback of tasks reduces the use of CPU compute time, since we reduce the duplication of work (see Fig. \ref{fig:processing-time}).
This is \textit{always} beneficial for utilisation, and energy efficiency, of computation.
However, reduced rollback does not have to reduce overall execution time.
In our experiments, overall execution time was indeed reduced consistently for the entire range of checkpoint levels.
The results are shown in Fig. \ref{fig:runtimes}, and use the same settings described in previous section.
We gain 1\% in overall execution time with checkpoint level 1 (most frequent), and 10\% with checkpoint level 6 (least frequent tested).
Overall, checkpointing as frequently as possible is recommended for 1D stencils.
This is in part due to the very efficient in-memory checkpointing using a peer node (e.g. Gamell et. al. \cite{Gamell2014} conclude that an MPI stencil code is optimally checkpointed every 4 iteration steps even for supercomputer runs). 
Another reason for our minimal checkpoint duration is the small amount of data to checkpoint for the one-dimensional case. A higher stencil dimensions would produce more interesting results, but is harder to illustrate and implement.

\begin{figure}
\centering
\begin{minipage}{0.5\textwidth}
\includegraphics[width=0.9\textwidth]{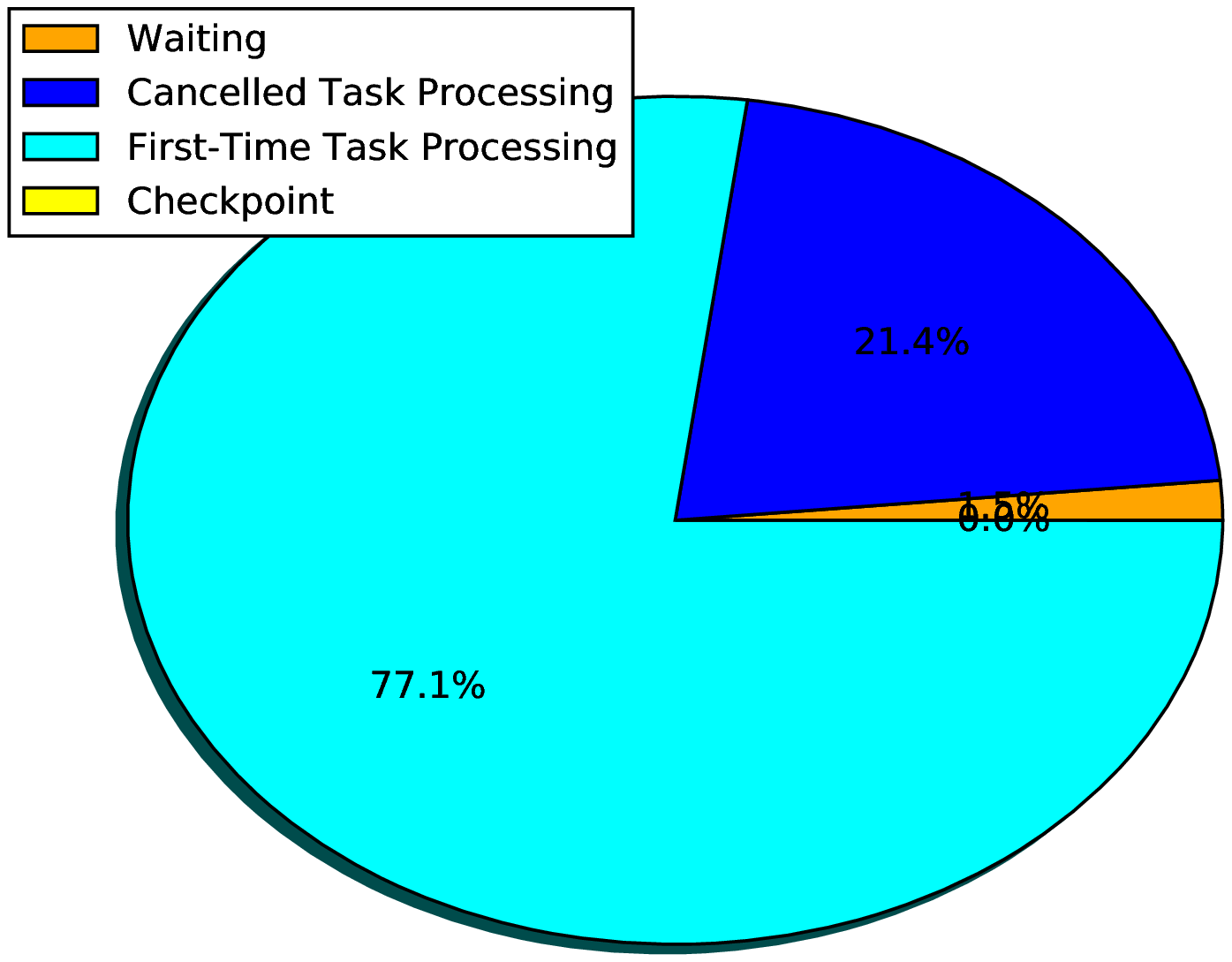}
\end{minipage}
\begin{minipage}{0.5\textwidth}
\includegraphics[width=0.9\textwidth]{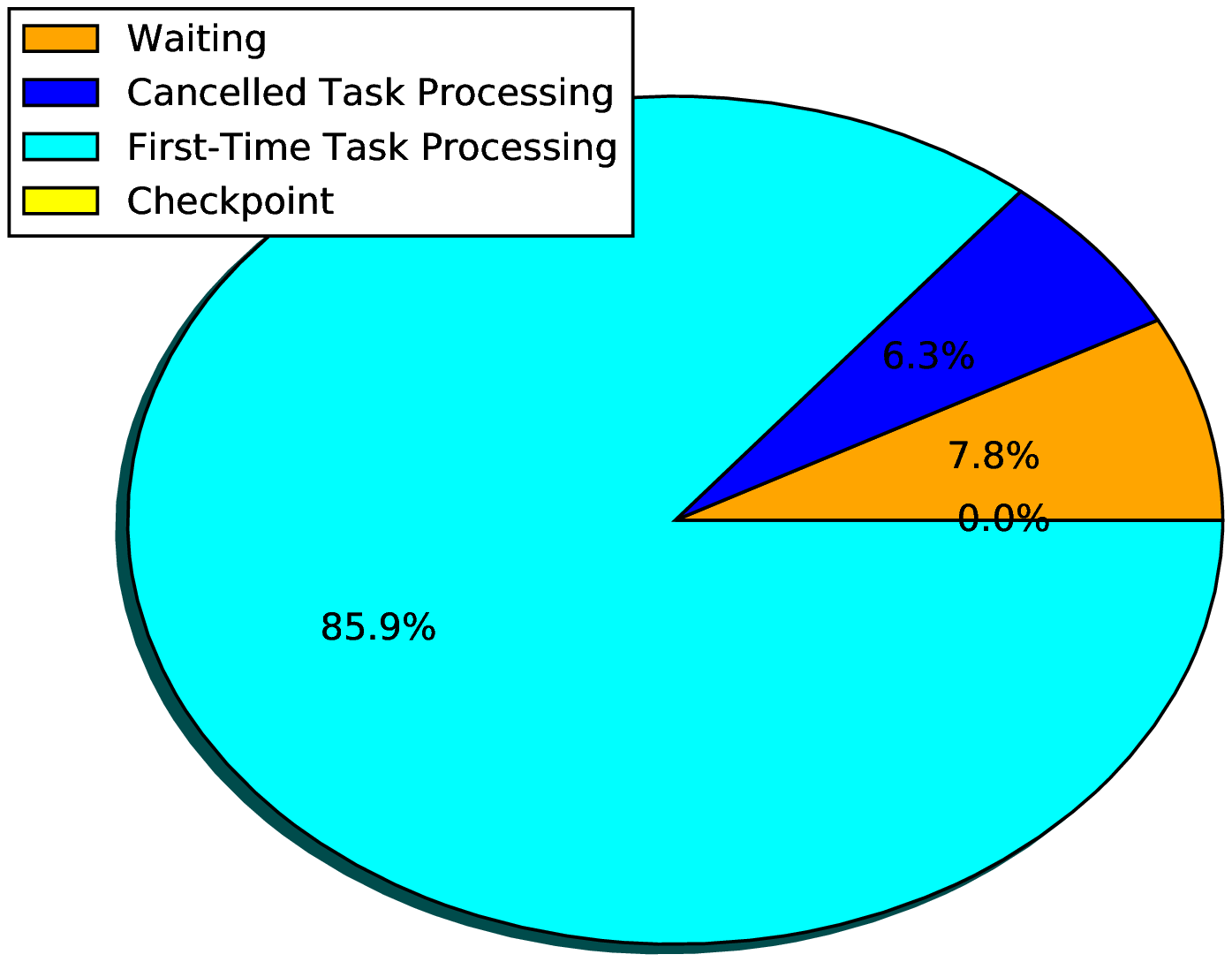}
\end{minipage}
\caption{Pie charts profiling a distributed execution with failures. $256^2$ $T_L$-level tasks, 5 random node failures in each run. Above: Default rollback. Below: Dependency-aware rollback.}
\label{fig:piecharts}
\end{figure}

We also profiled simulation runs to understand where time is spent throughout an execution for each of the implemented rollback strategies.
We used the previous setting of $256^2$ $T_L$-level tasks, and we take a snapshot for checkpoint level 6, and 5 total node failures, for a default rollback, and a dependency-aware rollback.
We enclose the results in Fig. \ref{fig:piecharts} as pie charts.
The different phases are aggregated from all workers.
As expected, processing tasks take the majority of execution time across both cases.
For both cases, checkpointing takes close to 0\%, due to the small size of checkpointed array (1.6MB), and its very efficient checkpointing at its guard node.
The interesting results are the time spent in processing cancelled and recomputed tasks, and the time spent waiting on tasks.
As confirmed in previous sections, dependency-aware rollback spends only a fraction of the time in processing cancelled tasks, because there are fewer of them.
The interesting finding here is that dependency-aware rollback spends a lot more time waiting on tasks to complete.
The reader may consider that each cancelled task may have one of two opposing effects:
\begin{itemize}
\item The recomputing of any already computed task may deteriorate overall runtime
\item However, the cancellation  and recomputing of a task, while wasteful, may improve overall runtime. Instead of being idle, a processor may then accelerate the recompute of failed tasks
\end{itemize}
The last observation explains why more time is wasted in waiting on incomplete tasks for dependency-aware rollback. 

\section{Discussion on Young/Daly formula}
\label{sec:discussions}
%\subsection{Applicability for General Kernels}
%
%A strong positive and general aspect of our work is our reduced rollback scheme, which holds for any application kernels.
%It is based on task dependencies; the kernel simply manifests one task dependency pattern.
%We are confident that this reduced rollback can easily be applied to other kernels, such as iterative map/reduce \cite{Ekanayake2010}, with beneficial performance.
%However, the checkpointing policy, and the lack of global communicators, needs to be reconsidered on a per-kernel basis.
%First, it is unrealistic to generalize checkpointing schemes, since they depend on the application data access patterns.
%Second, we maintain in Sect. \ref{sec:recovery} that a global communicator is not needed, and that supports the efficient recovery in step 1.
%Indeed, stencil kernels require no global communicator per se, since they rely on neighbor communication and not on collective communication.
%However, kernels with collective communication patterns may indeed require global communicators for performance reasons.

%\subsection{Young/Daly Formula}
%\label{sec:young-daly}
% Daly's formula for coordinated checkpointing has been extended in \cite{Bosilca2013} to include a slowdown factor as follows:
%\begin{equation}
%    T_{c} = \sqrt{2 * C * \mu * (1 - \alpha)}
%\end{equation}
%The parameter $\mu$ denotes the MTBF, and $C$ the checkpoint duration, $\alpha$ is the slowdown factor of checkpointing.
We performed extended experiments with varying checkpoint levels in our evaluation.
This may seem unnecessary to a fault tolerance expert in MPI applications; after all, the Young/Daly formula \cite{Young1974,Daly2006} is well established for finding the optimal checkpointing interval.
It has been verified, both theoretically and practically, to provide an optimal interval when running well structured MPI codes with traditional coordinated checkpointing approaches.

Unfortunately, there are a number of issues when applying this formula for dynamic task-based runtimes.
It is derived based on some assumptions:
\begin{itemize}
    \item The checkpoint duration and checkpoint intervals are assumed to be constant throughout the execution.
    \item In his study of optimal restart intervals, Daly assumes rollback to the last consistent checkpoint.
\end{itemize}
Unfortunately, none of these assumptions holds.
The former assumption is generally not true for dynamic task-based runtimes.
For example, we have illustrated the variable checkpointing duration in Fig. \ref{fig:two-schedules} for two marginally different schedules of task processing.
A large number of policies are sensible for task-based runtimes, potentially resulting in varying duration of both processing and checkpointing of tasks.
On the other hand, the latter assumption is not true for dependency-aware rollback, which does not restart all tasks from the last global checkpoint.
%The duration of checkpointing depends on the scheduling policy.
%Any application kernel (like stencil kernels) can clearly specify its predefined task tree and task dependencies.
%It is entirely up to the runtime scheduling policy how tasks will be processed.
It is an open question if Young/Daly formula is accurate for dynamic task-based runtimes.
We do not answer this question in this work.

\section{Conclusion and Future Work}
\label{sec:conclusion}
In this work, we introduced task logging into task-based runtimes, and integrated it into state-of-the-art checkpoint/restart mechanisms.
This allowed us to propose a dependency-aware rollback mechanism, which cancels a significantly smaller number of tasks during recovery from node failures.
The proposed reduced rollback strategy translated, as expected, into proportionally less computed tasks. 
For the kernel runs and 4 to 5 node failures per run, we gain between 1.3\% (very frequent checkpointing) and 16\% (infrequent checkpointing) in aggregated processing time.
Somewhat less intuitive is the fact that the overall runtime, when employing dependency-aware recovery, was consistently  faster than in the case of default rollbacks, with overall time reduction between 1\% and 10\%, once again depending on the frequency of checkpointing.
This is a surprising positive result, since we naturally observe that reduced rollback leads to more tasks being idle during execution.
To demonstrate the dependency-aware rollback, we used a stencil kernel, and implemented an intelligent and fine-grained checkpoint policy for this kernel.
We implemented the entire protocol in a simulator, which provided our evaluation.

The bulk of our future work is in fully implementing the proposed resilience protocol within the asynchronous global address space runtime HPX \cite{Kaiser2016}.
This implies some extensions to the runtime, such as the support for annotated data during the recomputing of tasks, which does not interfere with global data.
We also plan to design an efficient implementation of Alg. \ref{alg2} in HPX; we believe application-specific optimisation can be explored per use case, reducing the overhead of computing the set of restarted tasks.
We also plan to study how close Young/Daly formula is to the optimal checkpointing interval in the context of load balancing runtimes, such as HPX.
\section*{Acknowledgements}
We would like to thank Hans Vandierendonck for encouraging the exploration of dependency-aware rollback.

This work was partially supported by the AllScale project that has received funding from the European Union’s Horizon 2020 research and innovation programme under Grant Agreement No. 671603.

\bibliographystyle{IEEEtran}
\bibliography{references}
\end{document}